\documentclass[pre,preprint,showpacs,showkeys]{revtex4}
\usepackage{graphicx}
\usepackage{subfigure}

\begin{document}

\title{Controlling mechanical properties of a polymeric glass}

\author{George J. Papakonstantopoulos$^1$\footnote{Author to whom correspondence should be addressed. Electronic mail:gjpapakonsta@gmail.com}}
\author{Juan J. de Pablo$^1$}
\affiliation{$^1$Department of Chemical and Biological Engineering,
University of Wisconsin-Madison, Madison, Wisconsin, 53706, USA}

\date{\today}

\begin{abstract}
In this work we use molecular simulations to examine methods of controlling
mechanical properties of polymeric glass materials such as elastic moduli,
mechanical heterogeneity as well as their glass transition temperature.
We study filled and unfilled polymers and examine the effect of particle size, 
volume fraction and polymer-particle interactions. We identify a relationship 
between mobility and dynamic heterogeneity with elastic moduli and glass transition temperature.
\end{abstract}


\maketitle 

\section{Introduction}
A series of experiments have recently found evidence of dynamical
heterogeneities~\cite{ediger,spiess,sillescu} in amorphous glassy 
systems. Additionally, recent, simulation studies have proposed the 
existence of mechanical heterogeneities in such 
materials~\cite{BarratPRB05,yioryos,yioryos2,yioryos3}. 
Polymeric glasses have been found to exhibit domains of several
nanometers whose elastic moduli can vary appreciably~\cite{KenjiPRL2,bohme2}.
These mechanical heterogeneities have been found to be connected to the failure of 
these materials upon deformation~\cite{yioryos2,yioryos3}. It is obvious that 
it is of great importance to identify methods of controlling these mechanical
heterogeneities, especially when the material is used for the creation of nanostructures.

Particulate fillers are used extensively in the polymer industry to alter
material properties of polymeric material. The addition of particles
can lead to the strengthening of polymers extending considerably
their range of applicability. Understanding the molecular mechanisms 
that lead to such modifications in the properties of nanocomposites 
is essential for the design of materials that have a desirable behavior.

Polymer nanocomposites can exhibit an increase on the values of
the mechanical properties~\cite{vollenberg,ou,berriot} or a
decrease~\cite{schandler,narayanan}, depending on the nature of
nanoparticle-polymer specific interactions. Some of the 
parameters that affect the properties of a composite are the 
particle size, the volume fraction and the extent 
of agglomeration. Experiments suggest that as the size of the particle 
decreases, the changes in mechanical properties become more
pronounced~\cite{schandler,vollenberg,vieweg}.

Simulations provide a valuable tool for the study of nanocomposite
systems. They can offer useful insights into the spatial and structural
arrangement of the particles in the polymer matrix. In this work, we use
uniaxial deformations on a three-dimensional amorphous polymeric system. 
We show that, consistent with previous calculations, nanoparticles can
be used to alter the elastic modulus and the glass transition of a material. 
Additionally, the mechanical heterogeneity of polymers can be controlled
with addition of nanoparticles. Finally, we show that a relationship exists
between mobility and ``dynamic heterogeneity'' with elastic moduli and glass
transition.

\section{Methodology}
\subsection{Model and Simulations methodology}
In our simulations, the segments of the polymer molecules interact
via a pairwise, 12-6 Lennard-Jones truncated potential energy
function, shifted at the cutoff $r_{c}=2.5\sigma$,
\begin{equation}
U_{nb}(r)= \left\{ \begin{array}{ll} 4\varepsilon\left[(\frac{\sigma}{r})^{12}-(\frac{\sigma}{r})^{6} \right] - U_{LJ}(r_{c}),  & r\leq r_{c} \\
 \\
 0,   &r>r_{c} \end{array} \right.
\end{equation}
where $\varepsilon$ and $\sigma$ are the Lennard-Jones parameters
for energy and length, respectively, and $r$ is the distance between
two interaction sites. The bonding energy between two
consecutive monomers in the same chain is given by
\begin{equation}
U_{b}(r)=k\left( r-\sigma\right)^{2},
\end{equation}
with bond constant $k = 10^{3}\varepsilon/\sigma^2$.
Nanoparticle-polymer segment interactions are described through a
potential energy of the form~\cite{doxastakis}
\begin{equation}
 U_{nb}^{f}(r)= \left\{ \begin{array}{ll}
4\varepsilon_{f}\left[(\frac{\sigma_{f}}{r-R_{f}})^{12}-(\frac{\sigma_{f}}{r-R_{f}})^{6} \right] - U_{LJ}(r_{c}),   & r-R_{f}\leq r_{c}\nonumber\\
\\
0,   & r-R_{f}>r_{c}
\end{array} \right.
\end{equation}
where $R_{f}$ is the radius of a nanoparticle and
$\sigma_{f}=\sigma$. The polymer-particle types of interactions considered,
are ``strongly attractive'' ($\varepsilon_{f}=10\varepsilon$ and $r_{c}=2.5\sigma$),
are ``attractive'' ($\varepsilon_{f}=5\varepsilon$ and $r_{c}=2.5\sigma$),
and ``neutral'' ($\varepsilon_{f}=\varepsilon$ and $r_{c}=2.5\sigma$). 
As can be seen from the form of the potential in Eq. 3, the interaction between a particle and a
monomer is taken at a distance $R_{f}$ (the radius of the filler)
from the center of the particle. In that way the interactions
are accounted for from the surface of the nanofiller.

The systems considered in this work consist of, on average, 450 chains of $N = 32$ beads. 
Simulations are performed in the $NPT$ ensemble. The pressure in all simulations is kept 
constant at $P$ = 0.0. All quantities are reported in LJ units, reduced with respect to 
the monomer $\sigma$ and $\epsilon$. All configurations are equilibrated at a temperature 
$T$ = 1.2. Then the configurations are cooled down to $T$ = 0.0001 for a period of 1.2 million
steps. The timestep is $\delta t$=0.001. At the temperature $T$ = 0.0001 we perform an NVT 
calculation of $200000$ steps. We calculate a local mean-square displacement of each atom 
around its average position which is a measure of mobility. 

A Monte Carlo method is used to equilibrate the systems of interest in
this work. Trial particle displacements include random monomer and
nanoparticle translations. Reptation moves are also attempted,
implemented within a configurational-bias scheme to increase
performance~\cite{depablo2}. To further enhance sampling we implement 
double-bridging trial moves~\cite{depablo,theodorou2}. These moves
consist of a simultaneous exchange of distinct parts of two neighboring
chains, and are highly effective for configurational sampling of long
chain molecules. Double bridging allows for effective equilibration of
the systems considered in this work~\cite{depablo}. It is particularly
important in nanoparticle-reinforced polymers, where sampling the
correct structure and arrangement of long chain molecules around
nanoparticles can be particularly demanding. 

\subsection{Non-affine displacement field}
In this section we describe the technique used to calculate the non-affine displacement field 
that arises when a material is deformed. The energy of a configuration created after cooling 
down to $T=0.0001$ by NPT is first minimized. A uniaxial deformation is then applied of strain 
$\varepsilon_{ii}$, where $ii$ = $xx$ ,$yy$ or $zz$, by rescaling all the coordinates and the 
corresponding box length affinely. The energy of this affinely deformed configuration is minimized 
again, keeping the simulation box shape and volume constant. This process yields particle or 
segmental displacements relative to the affinely deformed state, the so called non-affine displacement 
field, $u(r)$. A conjugate gradient minimization technique is used. 

\section{Results and Discussion}
\begin{figure}[htp]
\begin{center}
\includegraphics*[width=7cm,angle=0]{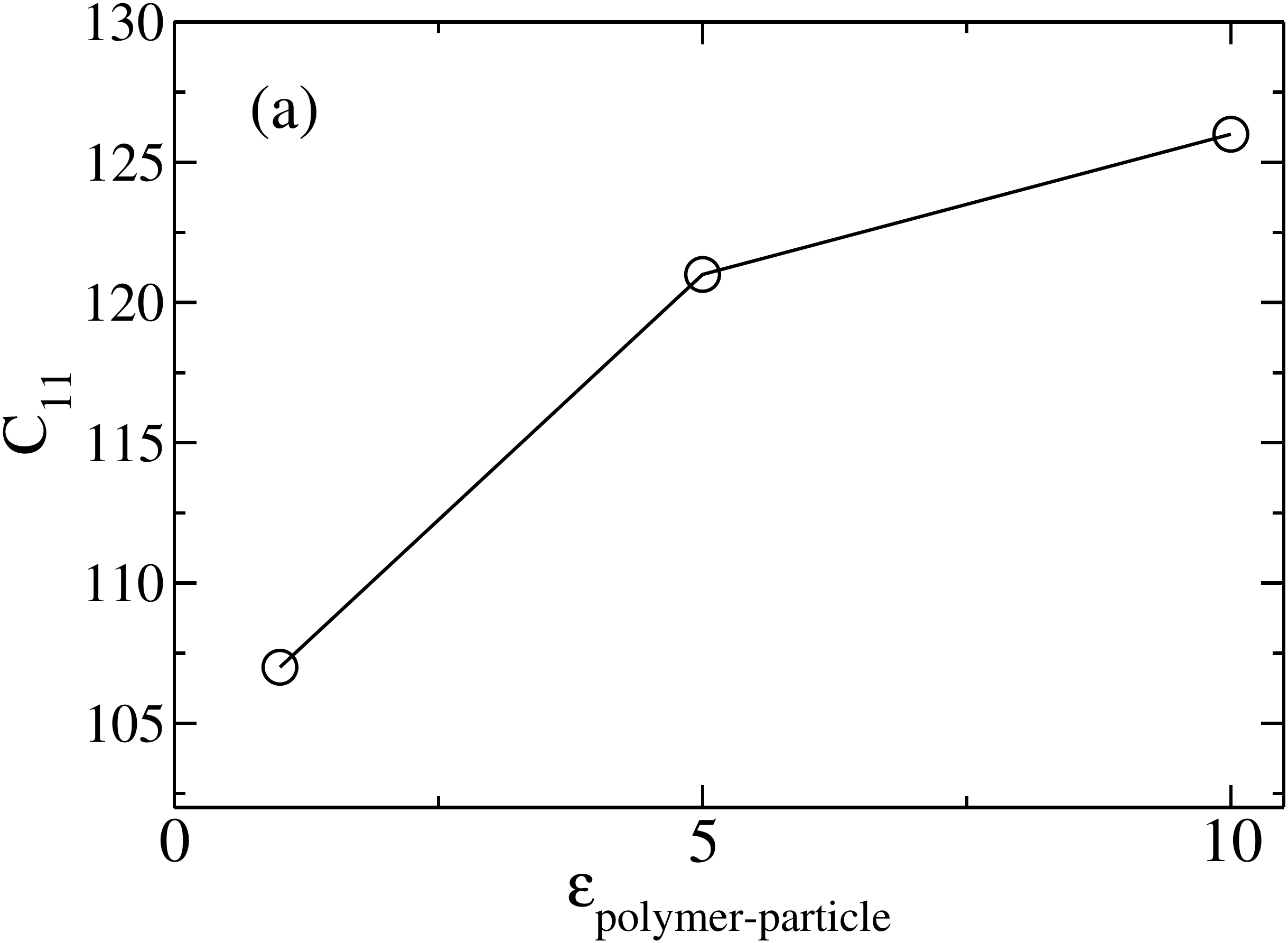}
\includegraphics*[width=7cm,angle=0]{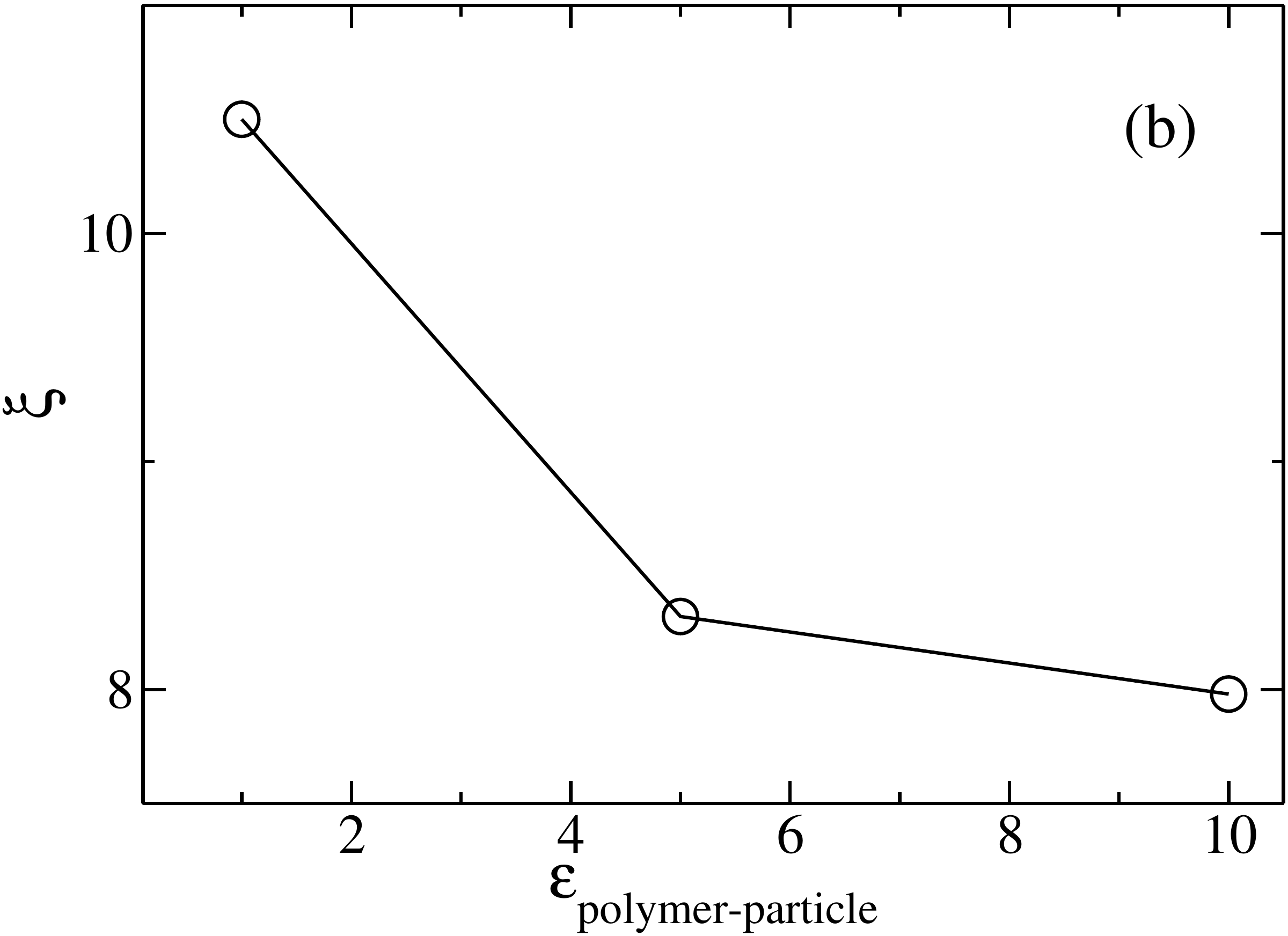}
\caption{a) Longitudinal elastic modulus as a function of particle-polymer interaction.
b) Non-affine displacement correlation length as a function of particle-polymer interaction.
} \label{figure:C11}
\end{center}
\end{figure}

We examine the effect of the particle-polymer interactions on the elastic longitudinal modulus
of the nanocomposites. As seen in Figure \ref{figure:C11}a for a system of $10\%$ volume fraction
and $R_{f}=2$, we find that increasing the 
attraction between polymer segments and nanoparticles results in an increase to the elastic 
modulus, $C_{11}$, in agreement with literature findings~\cite{yioryos,yioryos2}. We examine how the 
addition of nanoparticles affects the mechanical heterogeneity of the material.
A measure of the mechanical heterogeneity can be obtained by analyzing the 
non-affine displacements~\cite{leonforte}. The correlation of the non-affine 
field is analyzed by calculating the function
\begin{eqnarray}
C(r) = \frac{<u(r)\cdot u(0)>}{<u(0)^2>}
\end{eqnarray}
for all segments separated by a distance $r$. The correlation
length $\xi$ that arises from this relation is related to the fragility 
and mechanical stability of a material~\cite{yioryos2,silbert}.
We find that the correlation length decreases by increasing the polymer-particle
interaction (Figure \ref{figure:C11}b). This suggests that the nanocomposite system becomes more mechanically
homogeneous and thus more mechanically stable by increasing polymer-particle interaction.

\begin{figure}[htp]
\begin{center}
\includegraphics*[width=7cm,angle=0]{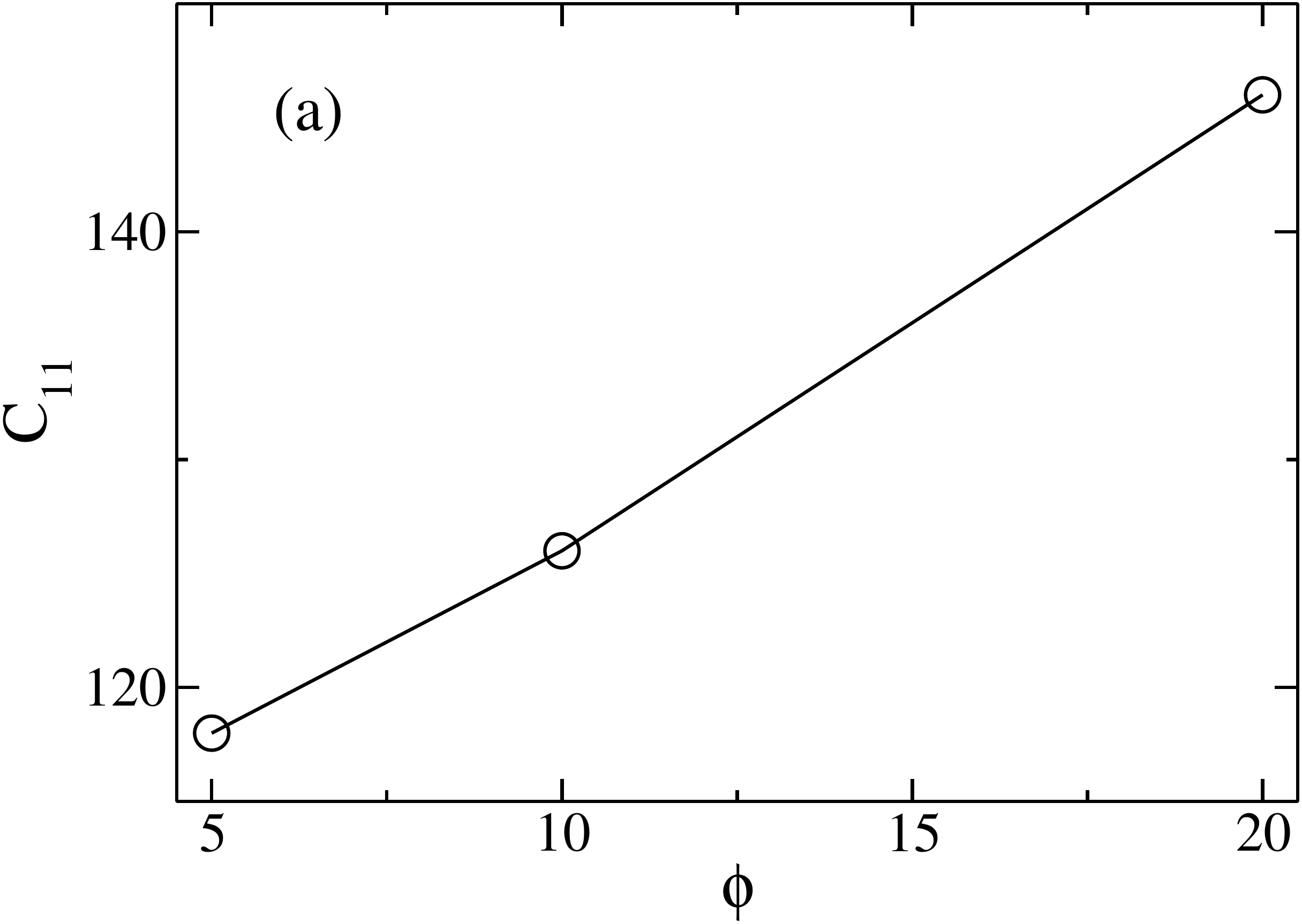}
\includegraphics*[width=7cm,angle=0]{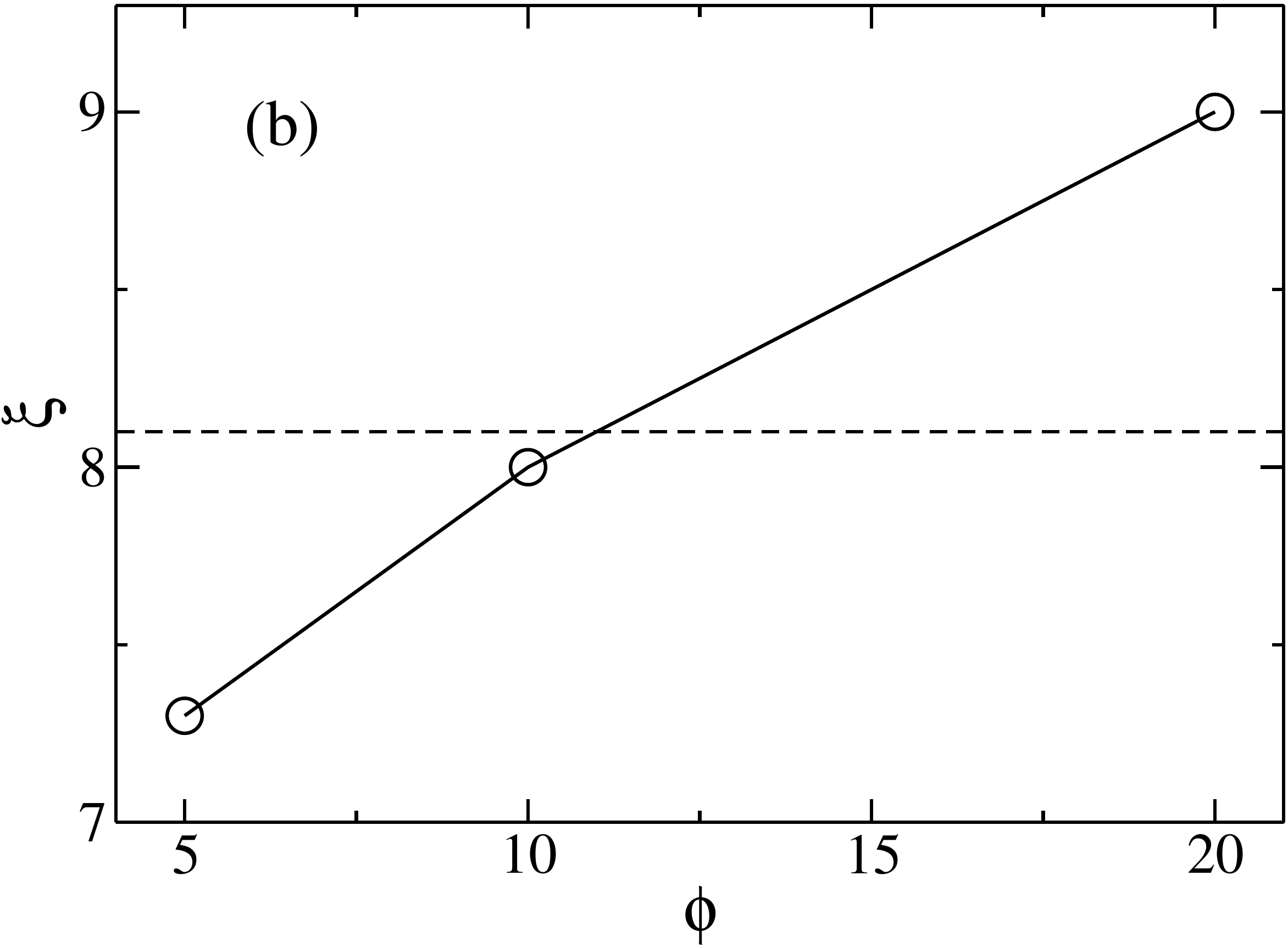}
\caption{a) Longitudinal elastic modulus as a function of particle volume fraction.
b) Non-affine displacement correlation length as a function of particle volume fraction.
The dashed line indicates the correlation length of the unfilled polymer.
} \label{figure:FI}
\end{center}
\end{figure}

We continue our investigation with the effect of particle volume fraction on the mechanical properties.
The particle size for this study was kept at $R_{f}=2$ and the polymer-particle interaction
was chosen $\varepsilon=10$. We find that increasing the volume fraction of the particles in the nanocomposite
results in an increase of the elastic modulus (Figure \ref{figure:FI}a). For all particle concentrations studies the elastic
modulus was higher than the pure polymer. Now we focus on the mechanical heterogeneity of these
systems. We find that increasing particle volume fraction, increases the mechanical heterogeneity
of the glass. However, as seen from Figure \ref{figure:FI}b, for small particle volume fractions
the heterogeneity of the polymeric nanocomposite glass is smaller than for the pure polymer. 
This indicates that a below a certain filler volume fraction, the nanocomposite polymeric 
glass is stronger but also less susceptible to failure in comparison to the pure polymer.

\begin{figure}[htp]
\begin{center}
\includegraphics*[width=7cm,angle=0]{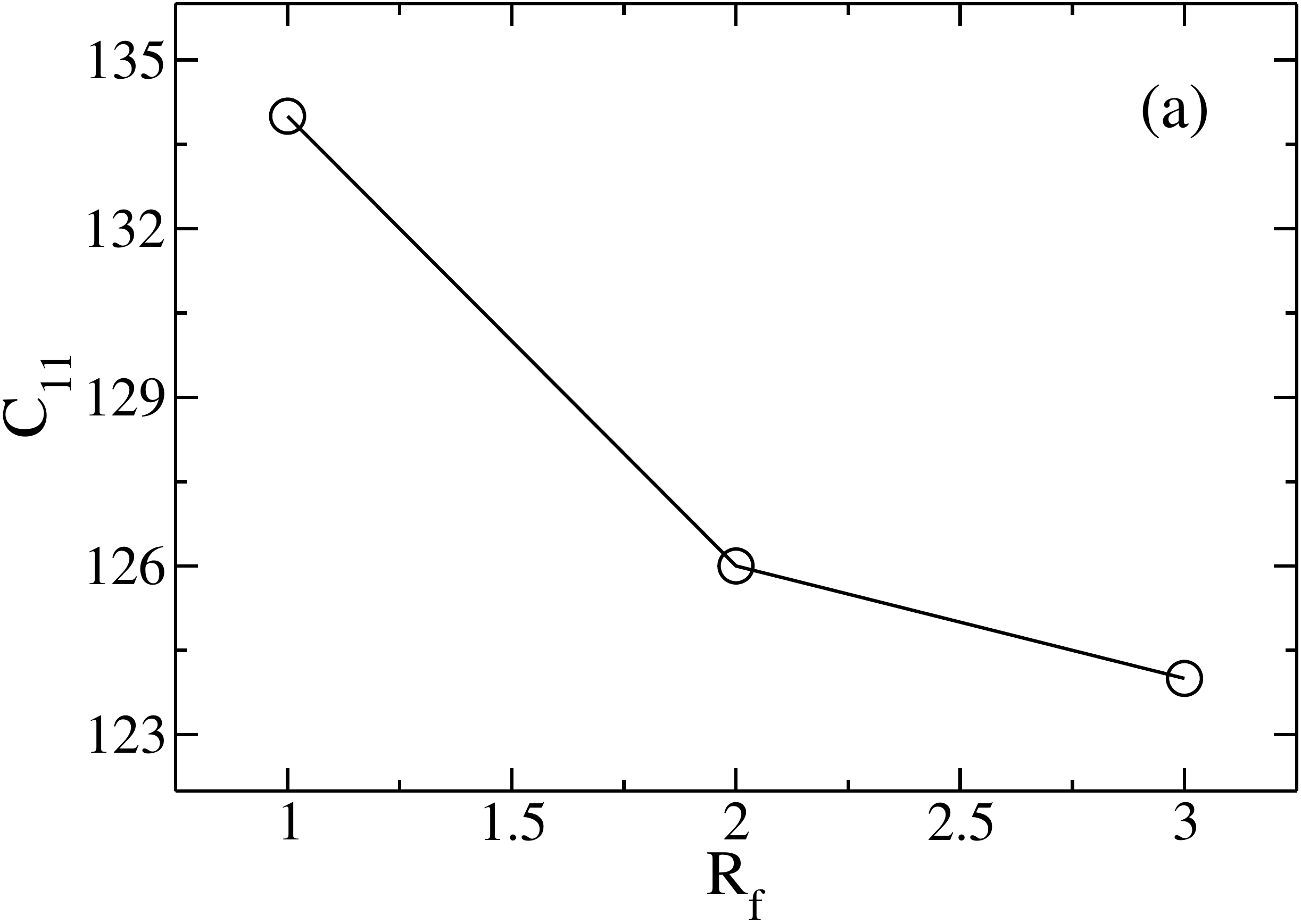}
\includegraphics*[width=7cm,angle=0]{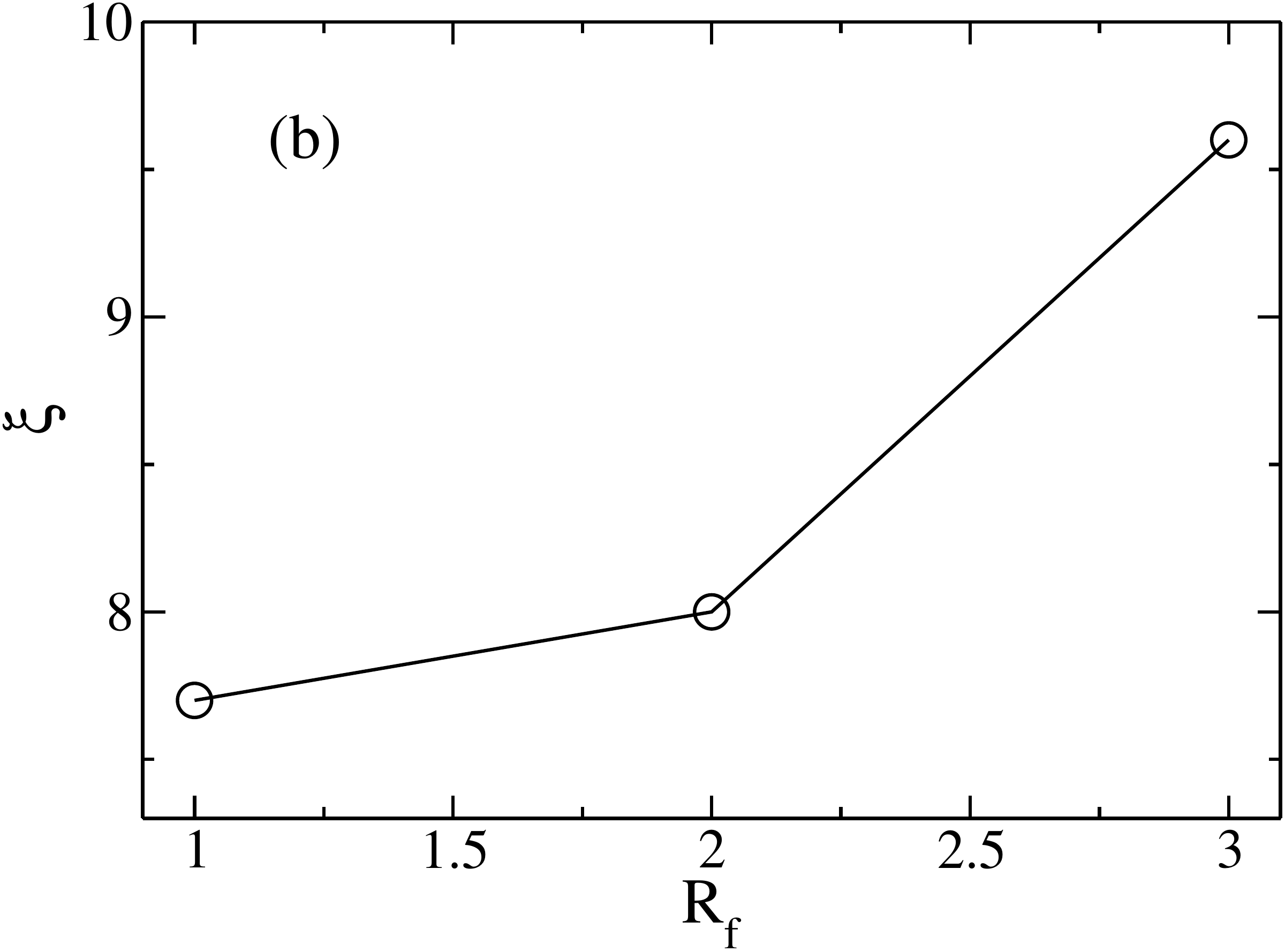}
\caption{a) Longitudinal elastic modulus as a function of particle size.
b) Non-affine displacement correlation length as a function of particle size
} \label{figure:RF}
\end{center}
\end{figure}

We now examine the effect of particle size on the mechanical properties.
As shown in Figure \ref{figure:RF}a, for a nanocomposite system of $10\%$ volume fraction
and polymer-particle interaction $\varepsilon=10$, decreasing particle size
results in a material of higher elastic modulus. This alteration of the
mechanical properties with particle size has been observed with experiments
where, depending on the polymer-particle interaction, the modulus
increases or decreases ~\cite{chahal,roberts,kopesky,mackay}.
What happens to the mechanical inhomogeneity of the material when the
particle size changes? In Figure \ref{figure:RF}b we plot the non-affine 
displacement correlation length as a function of particle size.
We see that decreasing particle size, the correlation length decreases
suggesting that the material becomes more mechanically stable. 
In addition, for the highest particle size studied the correlation length 
was found to be higher than the pure polymer while for the smallest particle
size it was significantly lower. 

\begin{figure}[htp]
\begin{center}
\includegraphics*[width=7cm,angle=0]{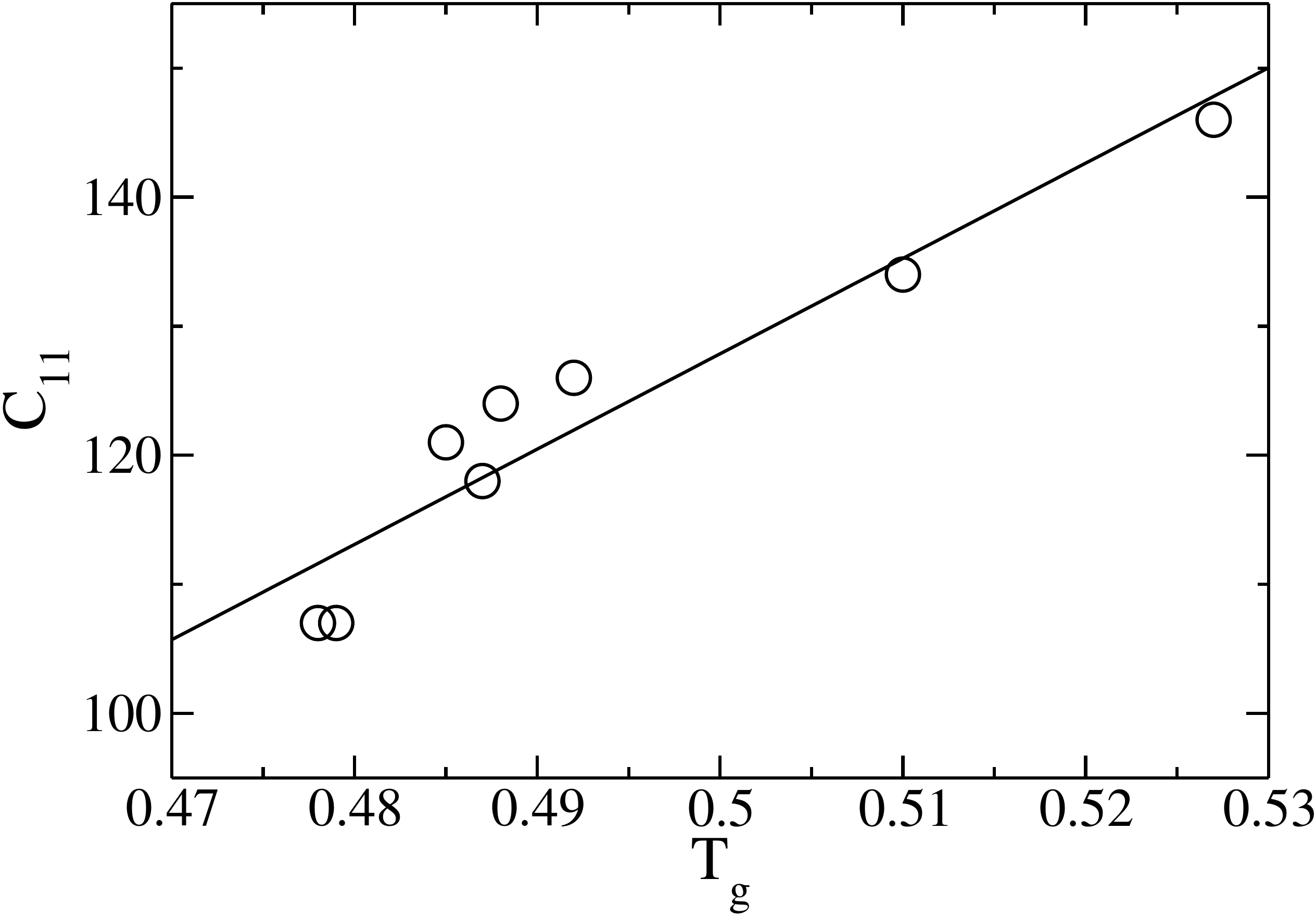}
\caption{Longitudinal elastic modulus as a function of glass transition temperature.
The line results from a linear regression to the data.
} \label{figure:C11TG}
\end{center}
\end{figure}

We continue with the examination of the relation of the glass transition temperature, $T_{g}$,
with the elastic modulus. We plot all the available data for all types of systems
used in this paper (different volume fraction, particle size, polymer-particle) 
in Figure \ref{figure:C11TG}.
We find that the higher the glass transition of the material studied,
the higher its elastic modulus which should logically be expected.

\begin{figure}[htp]
\begin{center}
\includegraphics*[width=7cm,angle=0]{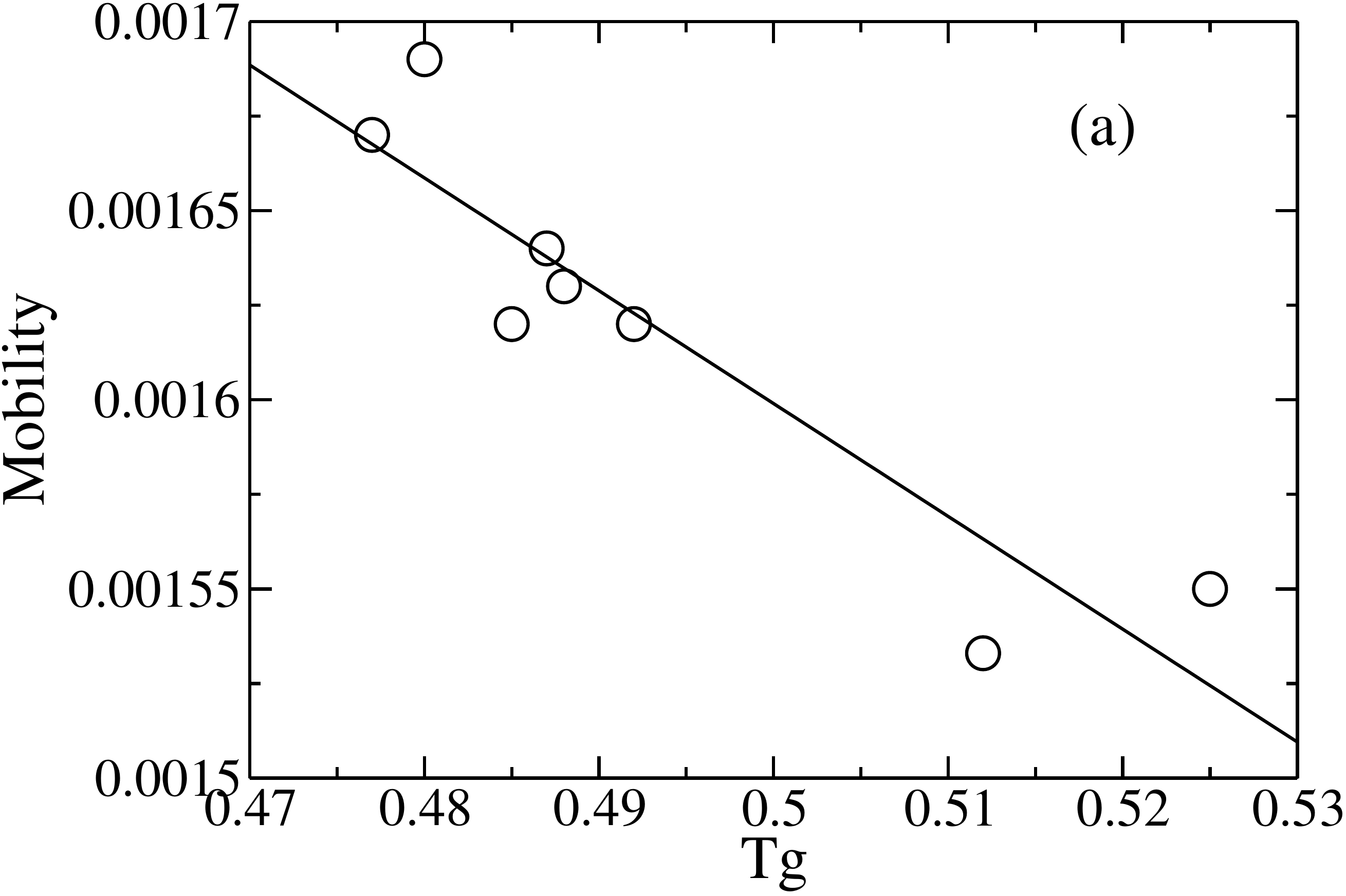}
\includegraphics*[width=7cm,angle=0]{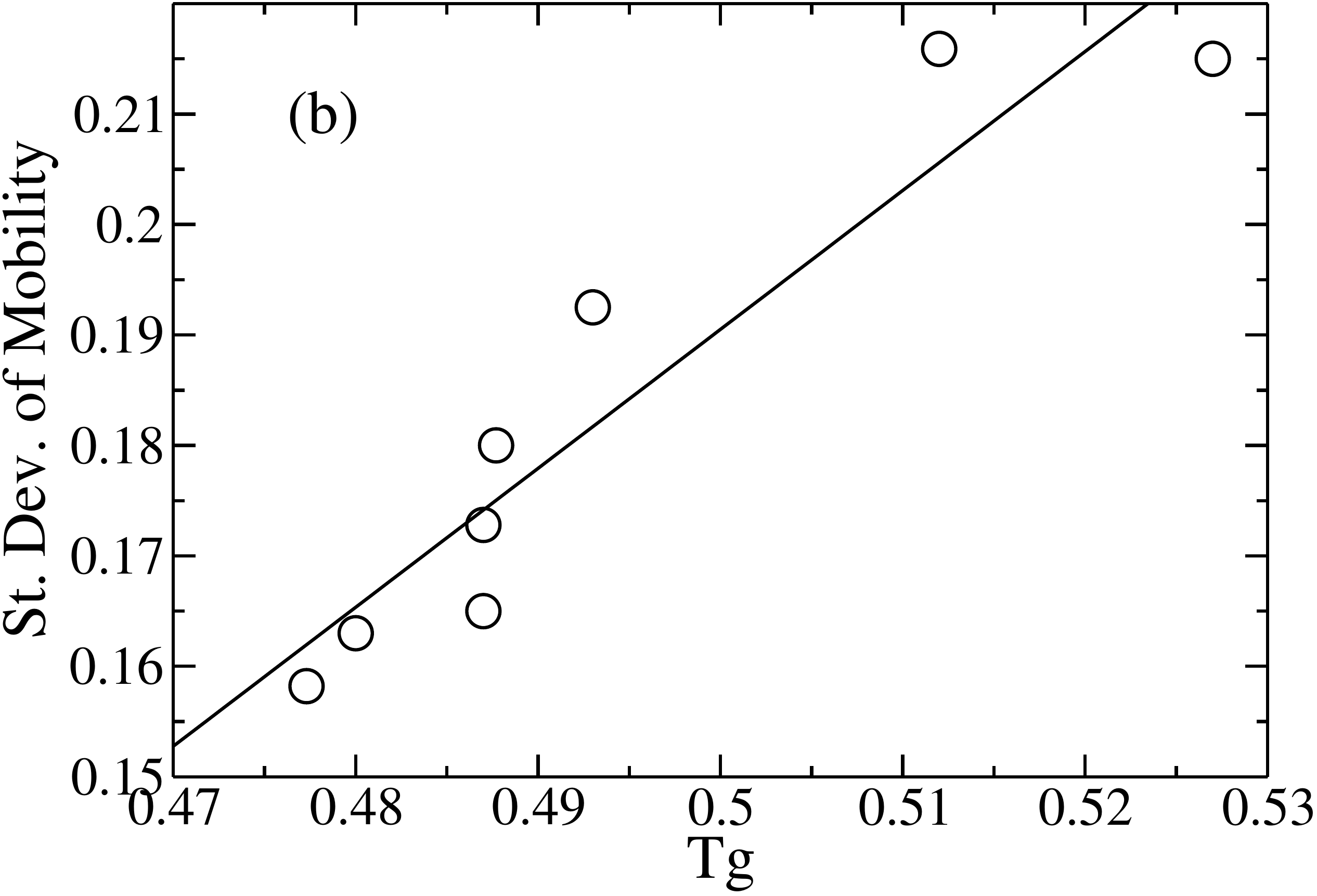}
\caption{Relationship of glass transition temperature and 
a) mobility and b) normalized fluctuations of mobility.
} \label{figure:TGDEB}
\end{center}
\end{figure}

But how is the mobility and the fluctuations of mobility, which are a measure of dynamic
heterogeneity, connected to $T_{g}$ and $C_{11}$? In Figure \ref{figure:TGDEB} we plot 
the mobility as a function of glass transition temperature. We find that materials of 
higher glass transition and elastic modulus exhibit a smaller mobility. Plotting the 
fluctuations of mobility with respect to the glass transition temperature in 
Figure \ref{figure:TGDEB} we see that the dynamic heterogeneities of a material exhibit 
the opposite behavior than the mobility with respect to the glass transition temperature 
and the elastic modulus. Materials of higher glass transition and elastic modulus exhibit 
a higher dynamic heterogeneity.

\begin{figure}[htp]
\begin{center}
\includegraphics*[width=7cm,angle=0]{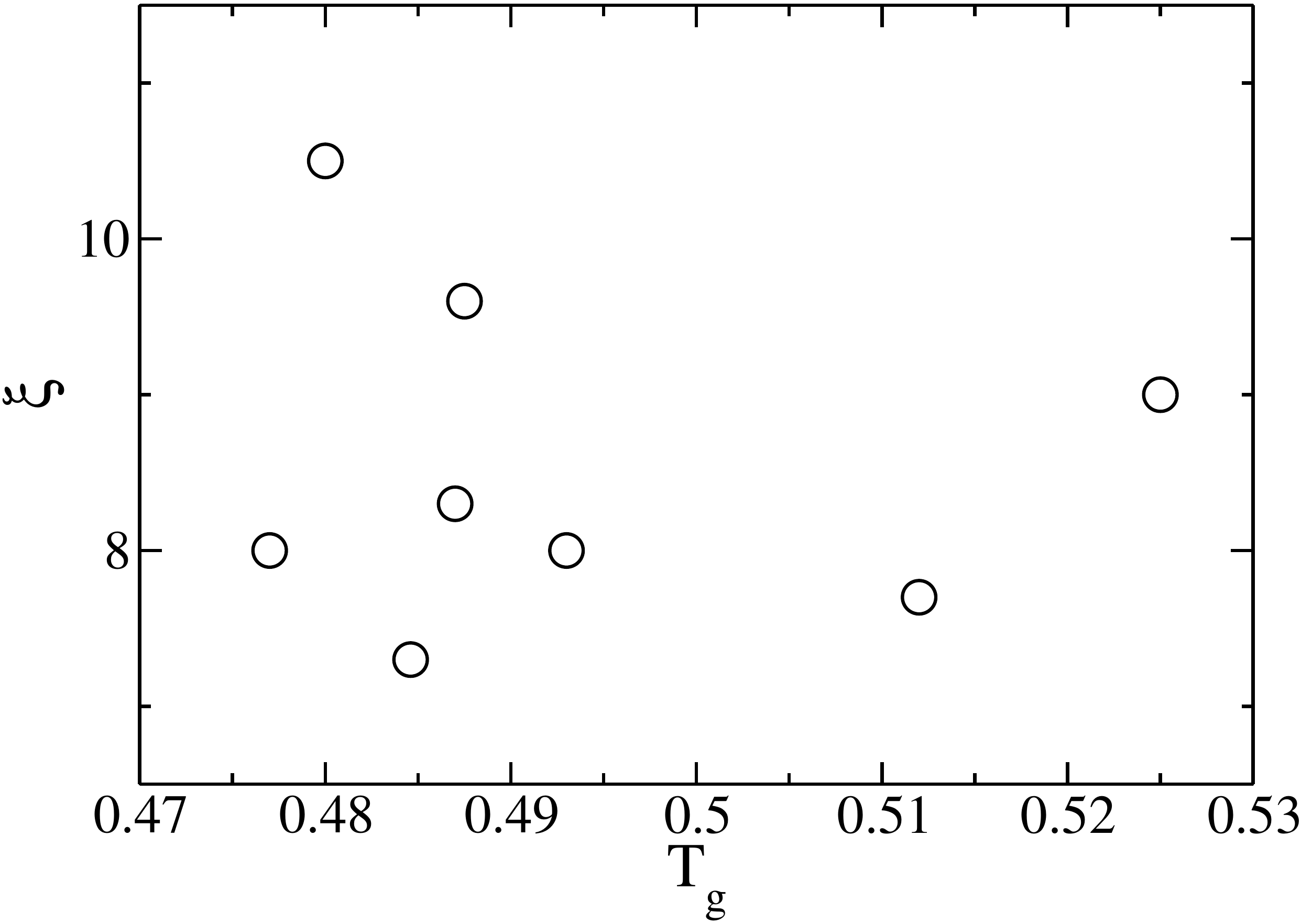}
\caption{Correlation length of non-affine displacements as a function of glass transition temperature.
} \label{figure:TGKSI}
\end{center}
\end{figure}

We now attempt to find a similar relation of the mechanical heterogeneities
to the elastic modulus and glass transition. We plot the correlation length
of the non-affine displacements with respect to the corresponding value of
$T_{g}$ in Figure \ref{figure:TGKSI}. The data appear to be scattered
and no clear connection is obvious for the mechanical inhomogeneity with $T_{g}$.

\section{Conclusions}
With the aid of molecular simulations we presented how the mechanical properties 
properties of a polymeric material can be controlled with the use of nanoparticle
inclusions. We studied the effect of polymer-particle interaction, particle volume
fraction and particle size on the elastic modulus and material failure. Additionally, 
we found a connection of glass transition temperature and elastic modulus with 
mobility and dynamic heterogeneities. No similar relation could be found between 
the mechanical heterogeneities and the elastic modulus. This work was supported by 
the NSF (NIRT Grand No. CTS-0506840) and the Semiconductor Research Corporation.

\bibliography{nc}

\end{document}